\def\year{2022}\relax
\documentclass[letterpaper]{article} 
\usepackage{aaai22}  
\usepackage{times}  
\usepackage{helvet}  
\usepackage{courier}  
\usepackage[hyphens]{url}  
\usepackage{graphicx} 
\usepackage{natbib}  
\usepackage{floatrow}%
\usepackage{float}
\usepackage{multirow}
\usepackage{verbatim}
\usepackage{placeins}
\usepackage{caption} 
\DeclareCaptionStyle{ruled}{labelfont=normalfont,labelsep=colon,strut=off} 
\usepackage{algorithm}
\usepackage{algorithmic}
\usepackage{amsmath}
\usepackage{cuted}
\usepackage{float}
\usepackage{xcolor}

\usepackage{newfloat}
\usepackage{listings}
\floatstyle{ruled}
\newfloat{listing}{tb}{lst}{}
\floatname{listing}{Listing}
\title{DeLoRes: Decorrelating Latent Spaces for Low-Resource\\ Audio Representation Learning}

\author{
    Sreyan Ghosh \textsuperscript{{\rm 1}}\equalcontrib,
    Ashish Seth \textsuperscript{{\rm 1}}\equalcontrib,
    Deepak Mittal \textsuperscript{{\rm 2}},
    Maneesh Singh \textsuperscript{{\rm 2}},
    S. Umesh \textsuperscript{{\rm 1}}
    \\
}

\affiliations{
    \textsuperscript{\rm 1}Speech Lab, Department of Electrical Engineering, Indian Institute of Technology Madras\\
    \textsuperscript{\rm 2}Verisk Analytics\\
    \ gsreyan@gmail.com, cs20s030@smail.iitm.ac.in,deepak.mittal@verisk.com, maneesh.singh@verisk.com,umeshs@ee.iitm.ac.in\\\

}

\usepackage{bibentry}

\begin{document}

\maketitle

\begin{abstract}
Inspired by the recent progress in self-supervised
learning for computer vision, in this paper we introduce \textbf{DeLoRes}, a new general-purpose audio representation learning approach. Our main objective is to make our network learn representations in a resource-constrained setting (both data and compute), that can generalize well across a diverse set of downstream tasks. Inspired from the Barlow Twins objective function, we propose to learn embeddings that are invariant to distortions of an input audio sample, while making sure that they contain
non-redundant information about the sample. To achieve this, we measure the cross-correlation matrix between the outputs of two identical networks fed with distorted versions of an audio segment sampled from an audio file and make it as close to the identity matrix as possible. We use a combination of a small subset of the large-scale AudioSet dataset and FSD50K for self-supervised learning and are able to learn with less than half the parameters compared to state-of-the-art algorithms. For evaluation, we transfer these learned representations to 9 downstream classification tasks, including speech, music, and animal sounds, and show competitive results under different evaluation setups. In addition to being simple and intuitive, our pre-training algorithm is amenable to compute through its inherent nature of construction and does not require careful implementation details to avoid trivial or degenerate solutions. Furthermore, we conduct ablation studies on our results and make all our code and pre-trained models publicly available \footnote{https://github.com/Speech-Lab-IITM/DeLoRes}.

\end{abstract}
\section{Introduction}

\noindent In recent times, unsupervised representation learning, including self-supervised and semi-supervised learning has shown great success across different modalities, for example, text, vision, and speech. Although self-supervised learning (SSL) and semi-supervised learning have achieved great performance on Automatic Speech Recognition (ASR), limited attempts have been made for other speech and audio processing tasks such as speech emotion recognition, speaker identification, language identification, and acoustic scene identification.

Self-supervised learning in speech and audio aims towards learning representations that contain high-level information from acoustic signals which can be further used in diverse sets of downstream tasks. Model weights learned through self-supervision are either used as feature extractors under the linear evaluation protocol \cite{niizumi2021byol, ling2020decoar} or used together with transfer learning for end-to-end fine-tuning with an added prediction-head for the downstream task \cite{chen2020simple, baevski2020wav2vec,hsu2021hubert}. Features learned through self-supervised speech representation learning have already proven to outperform other low-level features such as filter-banks and mel-frequency cepstral coefficients (MFCCs).  

In the past, a limited amount of work has been done in learning general-purpose audio representations that can perform well across a diverse set of downstream tasks beyond speech recognition and the one's proposed suffers from shortcomings. For example, triplet-based objectives used in \cite{jansen2017unsupervised,shor_2020} rely on the mining of negative samples and the quality of learned features varies significantly with the sample generation scheme \cite{shor_2020}. Contrastive learning systems used in \citep{saeed2020contrastive} typically work online and rely on a large number of explicit pairwise feature comparisons, which are computationally challenging, given the fact that they
require large batches for mining negative samples. Moreover, \cite{baevski2020wav2vec} show that the quality of mined negative samples can affect the performance. On the other hand, the method proposed by \cite{niizumi2021byol} uses a \emph{momentum encoder}, where a moving-average network is used  to produce prediction targets for optimizing the MSE loss between two batches of augmented samples of the same audio segments. However, the symmetry-breaking network design has been found crucial for this approach. Moreover, \cite{chen2020exploring} also conclude that the asymmetry of the learning update, ``stop-gradient'', is critical to preventing trivial solutions in this type of learning. Adding to this, the requirement of a separate teacher network also doubles the number of parameters required for training.


Prior studies in this domain use large data sets and large architectures for SSL pre-training. However, as correctly pointed out by \cite{hannun2021history}, the main challenges with self-supervision are those of scale, and hence accessibility. SSL would therefore be more accessible given lighter-weight models which could be trained efficiently on fewer data. Additionally, another problem is that many of these methods utilize the time-series aspect of audio signals, i.e., audio segments cropped closer in time are expected to have closer representations, whereas those far away in time are expected to have distanced representations. However, as pointed out by \cite{niizumi2021byol}, using contradictory examples of music and gunshot, this cannot be deemed as the optimal strategy in all cases.

In this paper we propose, \textbf{DeLoRes}, \textbf{De}correlating Latent Spaces for \textbf{Lo}w \textbf{Res}ource Audio Representation Learning, a simple yet powerful self-supervised pre-training framework to learn general-purpose audio representations of sounds beyond and including speech. Inspired by the Barlow Twins framework \cite{zbontar2021barlow}, we employ an invariance and redundancy-reduction based objective function, where we try to make the cross-correlation matrix, computed from embeddings of a pair of augmented samples of the same audio segments, as close to the identity matrix as possible. This idea fits well to the acoustic domain wherein the cross-correlation measure is generally used to calculate the correlation between two signals shifted in time. Compared to other methods in the self-supervised audio pre-training domain, our methodology is conceptually simple and avoids trivial solutions by the inherent nature of its construction. We alleviate the requirement of large batches as in \cite{saeed2020contrastive} and avoid the reliance on symmetry-breaking network designs for distorted samples using an extra teacher network as in \cite{niizumi2021byol}, the requirement for carefully curation of a sample generation scheme for mining negatives as in \cite{jansen2017unsupervised}, or the possibility of degenerate solutions as in \cite{ghosh2021deep}.

We demonstrate the effectiveness of DeLoRes over 9 challenging and diverse downstream tasks including speech, music, and animal sounds. We pre-train DeLoRes on a small subset (10\%) of the large-scale AudioSet dataset \cite{gemmeke2017audio} with a much smaller architecture compared to prior art, and show that only a linear classifier trained over DeLoRes embeddings is competitive in performance to other state-of-the-art (SOTA) algorithms for learning general-purpose audio representations, using much lesser data and compute than their implementations. 



\section{Related Work}

Recently, SSL applied to on unlabelled data has proven to be very effective, achieving SOTA performance when used alongside transfer learning on various low-resource downstream tasks, in various modalities such as text, image, and speech. It has also achieved significant performance boosts or close to SOTA results under the linear evaluation protocol, which allows only a linear layer to be trained on top of the representations learned by SSL, for a particular downstream task.  

SSL algorithms, that have shown great success, include variations of contrastive learning, masked prediction, clustering, or the use of \emph{momentum encoders} optimized with diverse objective functions. A simple yet powerful idea is to compare pairs of image representations to push away representations from different images while pulling together those from transformations, or views, of the same image. \cite{chen2020simple} was the first to propose the use of this framework in CV and proposed to calculate a contrastive loss by taking different augmentations of the same image as positive pairs and other images in the batch as negatives. 

Recent \emph{non-contrastive} SSL methods include \cite{chen2020exploring} and \cite{grill2020bootstrap}, where both the network architecture and the parameter updates are modified to introduce asymmetry. The asymmetry in network architecture is created by using a specific ``predictor'' network and the parameter updates are asymmetric such that the model parameters are only updated using one distorted version of the input, while the representations from another distorted version are used as a fixed target.

In another line of work, clustering methods such as \cite{caron2019deep} propose computing ``pseudo-labels'' from one view and predicting the cluster assignment using another view of the same image. On similar lines, \citep{caron2021unsupervised} and \cite{asano2020selflabelling} propose the use of non-differential operators. \citep{caron2021unsupervised} draws its inspiration partly from contrastive learning and proposes a swapped cluster prediction problem, whereby they try to predict the cluster assignment of one view from the other and optimize their network using cross-entropy loss.


SSL solving the masked prediction task has been prevalent in the text domain in the form of Masked Language Modelling (MLM) and in speech as Masked Acoustic Modelling (MAM). Most of these approaches aim to either predict the class of the masked entity using a classification objective as in \cite{hsu2021hubert,devlin2019bert} or reconstruct the original frame as in \cite{liu2020mockingjay,liu2021tera} or to enforce similarity between the prediction of the network for the masked frame and a quantized representation of the original masked frame as in \cite{baevski2020wav2vec}.

Prior work on unsupervised audio representation domain is diverse. For example, L3 \cite{arand2017look}, AuDeep \cite{freitag2017audeep}, autoregressive predictive coding \cite{chung2020generative}, contrastive predictive coding \cite{oord2019representation}, metric
learning \cite{jansen2017unsupervised}, autoencoding \cite{latif2020variational}. However, these methods were
evaluated on just one or a limited set of downstream tasks. TRILL \cite{trill} or TRIpLet Loss network was one of the first works to test its learned representations on a diverse set of downstream tasks. TRILL optimizes a triplet-based objective function whereby the network represents audio such that segments that are closer in time are also closer in the embedding space. The anchor and positives were sampled from the same audio sample while the negatives were sampled from a different sample.

COLA \cite{saeed2020contrastive} solves a contrastive task for learning general-purpose audio representations, which outperforms prior art in this domain. They also employ a mining strategy similar to TRILL wherein they generate similar pairs by simply sampling segments from the same audio sample and negative samples from different audio samples. They demonstrate the effectiveness of two objective functions, namely cosine and bilinear similarities, and also show that larger batch sizes help in learning better representations.

On the other hand, BYOL-A \cite{niizumi2021byol} proposes a different approach in that it does not use negative samples. Instead, it directly minimizes the mean squared error of embeddings originating from the same audio segment with contrasts created by data augmentations. This approach also differs from prior-art in which they learn audio representations from a single audio segment without expecting relationships between different time segments of audio samples.

Very recently, DECAR \cite{ghosh2021deep} proposed a clustering framework to learn general-purpose audio representations, based on the Deepcluster framework in CV. Their pre-training approach is based on two forward passes through the network whereby they first cluster the output of the feature extractor network and then use the subsequent cluster assignments as ``pseudo-labels'' for an augmented version of the same audio sample. In their second pass through the network, the loss between the network prediction of the augmented sample and the pseudo-label from the first pass is used to optimize the network weights.

Inspired by the transformers, that were originally developed for text processing but have shown great success in image and speech domains too, \cite{gong2021ssast} proposed SSAST, a self-supervised pre-training strategy for the audio spectrogram transformer AST \cite{gong2021ast}. AST is the first convolution-free, purely attention-based model for audio classification. In SSAST, the authors optimize a joint
discriminative and generative masked spectrogram patch
modeling (MSPM) objective whereby they sum InfoNCE loss and mean-squared-error, where both of the losses are weighted by a hyper-parameter lambda.

While contrastive learning approaches need large batch sizes and careful mining of negative samples for training, one common problem with \emph{non-contrastive} SSL approaches in both CV and the audio domain, is that there exist trivial solutions to the learning objective that these methods avoid via particular implementation choices or they are the result of non-trivial learning dynamics adopted by them. Thus, the recently proposed Barlow Twins makes itself unique from prior art by its new loss function which is conceptually much simpler and avoids pitfalls through its inherent nature of construction. Similar to other approaches, the Barlow Twins also works on a pair of differently augmented views of the same image, however, now it calculates the cross-correlation matrix between the embeddings of the network which was fed with the batches of these augmented views, and tries to make this matrix as close to identity as possible. Our approach in this paper is inspired by the Barlow Twins framework.

\section{Methodology}

\subsection{DeLoRes Architecture}

For pre-training the SSL, we use a convolution-based feature encoder. The learned weights are then transferred to other downstream tasks. 
We experiment with several ConvNet architectures, including the highly-scalable EfficientNet-B0 which has been commonly used in prior work \cite{saeed2020contrastive,ghosh2021deep}. However, we achieved better results when we resorted to a simpler architecture which was submitted as a solution of Task 6 of the Detection and Classification of Acoustic Scenes and Events (DCASE) 2020 Challenge \cite{koizumi2020ntt,takeuchi2020effects}. Beyond being much simpler, the architecture has proven to perform reasonably well in a challenge related to the task we are trying to solve. This architecture has also been used in \cite{niizumi2021byol} for pre-training the SSL. 


The network architecture is composed of three Conv2D layers, each followed by a BatchNormalization layer, a rectified linear unit (ReLU) activation function, and a MaxPool2D layer. Finally, we pass the embeddings obtained through a set of linear layers, each followed by a ReLU activation function again and we also use dropout layer between the two layers. 

In the SSL pre-training stage, for the \emph{projection head}, instead of three linear layers as in \cite{zbontar2021barlow}, we use two, with the first one followed by BatchNorm1D and a ReLU activation function. Additionally, we employ a dropout layer between the projection head and encoder output to add regularisation and then pass the final embeddings through a BatchNormalization layer to make them zero mean and unit variance. The number of units in the projection head is a tunable parameter and affects the size of our cross-correlation matrix $\mathcal{C}$ as defined in equation (\ref{eqn:cross}).

As mentioned earlier and shown in Fig. 1, we use the projection head just for our SSL pre-training task and discard it after this step. More layer-specific details can be found in Table 1.

\subsection{DeLoRes Learning Algorithm}

\subsubsection{Co-variance and Cross-correlation}

Given two time series $x_t$ and $y_t$ of $N$ number of samples, where $t$ is the time index, and $x_{t-\tau}$ is the delayed version of $x_t$ by $\tau$ samples, The cross-covariance matrix between the pairs of signals can be calculated as:

\begin{equation}\label{eq: covar}
\sigma_{x y}(\tau)=\frac{1}{N-1} \sum_{t=1}^{N}\left(x_{t - \tau}-\mu_{x}\right)\left(y_{t}-\mu_{y}\right)
\end{equation}

where $\mu_x$ and $\mu_y$ are the means of $x_t$ and $y_t$, respectively.
By normalizing the equation (\ref{eq: covar}), the cross-correlation of the two signals can be calculated as:

\begin{equation}\label{eq: correlation}
\rho_{x y}(\tau)=\frac{\sigma_{x y}(\tau)}{\sqrt{\sigma_{x x}(0) \sigma_{y y}(0)}}
\end{equation}
 
\subsubsection{Objective Function}

We employ the Barlow Twins objective function \cite{zbontar2021barlow}, defined in equation (\ref{eqn:barlow}), to optimize our network.

\begin{table}[H]
    \centering
    \resizebox{8.7cm}{!}{%
    \begin{tabular}{|l|l|l|}
    
    \hline
    \textbf{Layer -- \#}    & \textbf{Layer Parameters} &         \textbf{Output} \\
    \hline
    Conv2D-1    & 3x3@64 &     [B, 64, 64, 96]  \\
    \hline
    BatchNorm2D-2    &     &     [B, 64, 64, 96]     \\
    \hline
    ReLU-3    &     &     [B, 64, 64, 96]     \\
    \hline
    MaxPool2D-4    & 2x2,stride=2 & [B, 64, 32, 48] \\
    \hline
    Conv2D-5    & 3x3@64 &     [B, 64, 32, 48]  \\
    \hline
    BatchNorm2D-6    &     &     [B, 64, 32, 48]     \\
    \hline
    ReLU-7    &     &     [B, 64, 32, 48]     \\
    \hline
    MaxPool2D-8    & 2x2,stride=2 & [B, 64, 16, 24] \\
    \hline
    Conv2D-9    & 3x3@64 &     [B, 64, 16, 24]  \\
    \hline
    BatchNorm2D-10    &     &     [B, 64, 16, 24]     \\
    \hline
    ReLU-11    &     &     [B, 64, 16, 24]    \\
    \hline
    MaxPool2D-12    & 2x2,stride=2 & [B, 64, 8, 12] \\
    \hline
    Reshape-13  &     &        [B, 12, 512]     \\
    \hline
    Linear-14    & out=2048 &   [B, 12, 2048]    \\
    \hline
    ReLU-15    &     &        [B, 12, 2048]    \\
    \hline
    Dropout-16    & 0.3 &        [B, 12, 2048]    \\
    \hline
    Linear-17    & out=2048 &   [B, 12, 2048]    \\
    \hline
    ReLU-18    &     &        [B, 12, 2048]    \\
    \hline
    $max(\cdot)\oplus mean(\cdot)$-19    &     &            [B, 2048]  \\
    \hline
    \end{tabular}%
    }
    \caption{Encoder network architecture}
    \label{table:1}
\end{table}

\begin{equation}
\label{eqn:barlow}
\mathcal{L} \mathcal{=} {\underbrace{\sum_{i}\left(1-\mathcal{C}_{i i}\right)^{2}}_{\text {Invariance Term }}}+ \lambda {\underbrace{\sum_{i} \sum_{j \neq i} \mathcal{C}_{i j}^{2}}_{\text {Redundancy Reduction Term }}}
\end{equation}
where $\mathcal{C}$ is defined in equation (\ref{eqn:cross}). The \emph{invariance term} of the objective function tries to make the  diagonal elements of the cross-correlation matrix close to 1, which intuitively makes the embedding invariant to the distortions applied. The \emph{redundancy reduction term}, which is weighted by a positive constant, $\lambda$, sums up the off-diagonal elements of the cross-correlation and tries to make it close to 0. Intuitively,  this decorrelation reduces the redundancy between output embeddings so that they contain non-redundant information about the sample and preserve only useful information.

The Barlow Twins objective function calculates the cross-correlation matrix between the two embeddings obtained from the encoder output for a pair of differently augmented versions of an audio sample and tries to make it as close to the identity matrix as possible.

\subsubsection{Learning Framework}
\begin{figure*}[tp]
  \centering
  \includegraphics[width=1\textwidth]{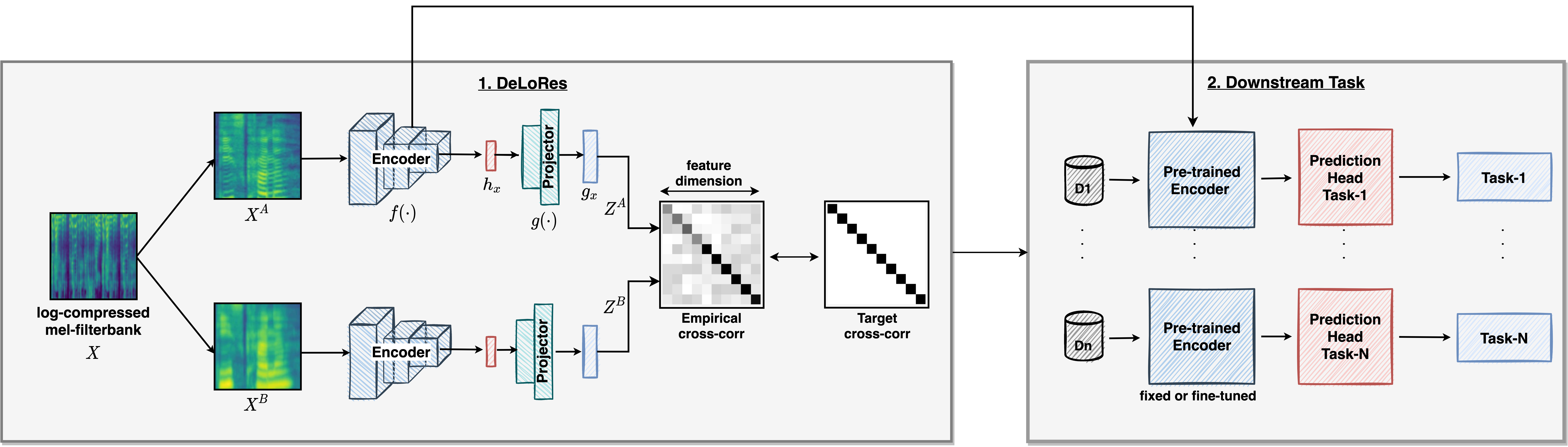}
  \caption{The block diagram of DeLoRes in pre-training and fine-tuning phases}
  \label{pic:barlow}
  \label{fig:lab1}
  \vspace{\floatsep}
\end{figure*}

Figure \ref{pic:barlow} shows the complete end-to-end pre-training and fine-tuning procedure. For a batch of audio samples, we first chunk a random segment $x$ of length 980 ms, from each audio sample, and then extract the log-compressed mel-filterbanks for each sample. Next, these samples are batched to obtain $X$, and we create exactly two augmented views $X\textsuperscript{A}$ and $X\textsuperscript{B}$, for each sample in the batch by applying different augmentations $K$. The ConvNet feature encoder $f$ is then used to obtain batch of embeddings $Z\textsuperscript{A}$ and $Z\textsuperscript{B}$. These outputs are then used to calculate $\mathcal{C}$, the cross-correlation matrix, where each element in the matrix is calculated by equation (\ref{eqn:cross}) along the batch dimension.

\begin{equation}
\label{eqn:cross}
\mathcal{C}_{i j} {=} \frac{\sum_{b} z_{b, i}^{A} z_{b, j}^{B}}{\sqrt{\sum_{b}\left(z_{b, i}^{A}\right)^{2}} \sqrt{\sum_{b}\left(z_{b, j}^{B}\right)^{2}}}
\end{equation}
where $b$ indexes batch samples, $i$ and $j$ indexes the vector dimension of the embeddings, and $\mathcal{C}$ can take values in a range -1 and 1.
While \cite{zbontar2021barlow} show how the design of the loss function was motivated by the Information Bottleneck theory, \cite{tsai2021note} have a great explanation why Barlow Twins can also be called \emph{negative-sample-free contrastive learning}. This explanation is beyond the scope of this paper and we would like our readers to refer to these papers directly.





\section{Data Sets}

We pre-train DeLoRes embeddings on a combination of a class-balanced subset of the large-scale Google AudioSet \cite{gemmeke2017audio} and the FSD50K dataset \cite{fonseca2020fsd50k}. The AudioSet subset consists of 0.2 million utterance or approx. 10\% of the total number of audio files in the original large-scale data set (2 million). We do this for two primary reasons:

\begin{itemize}
    \item All the AudioSet files are not readily available for download and need to be sliced manually from YouTube videos. Moreover, with time, a lot of these videos have been taken down by YouTube. Thus, results reported on pre-training on complete 2 million utterances of the AudioSet might not be reproducible by independent researchers.
    \item As pointed out by \cite{hannun2021history}, self-supervised learning in speech and audio would be more accessible to smaller labs given lighter-weight models which could be trained efficiently on a lesser amount of data. Thus, we wanted to compare the performance of our system on a smaller pre-training dataset setting, contrary to the prior art in this domain.
\end{itemize}

For the downstream tasks, we test the efficiency of our learned embeddings on a diverse set of audio tasks. We take a mixture of both speech and non-speech tasks to increase variability and test generalizability. 
For speaker identification we resort to two commonly used datasets, namely, LibriSpeech (LBS) \cite{7178964} and Voxceleb \cite{Nagrani_2017}. For the task of keyword spotting, we use the Google Speech Commands V1 (SC) and V2 datasets \cite{warden2018speech}. This dataset is used in various label settings in prior art. We use both the 12 and 35 label settings for our experiments. 
For speech classification, we also apply speech emotion classification to the IEMOCAP data set \cite{busso2008iemocap} and language identification on the Voxforge \cite{Voxforge.org} data set. For classifying acoustic signals beyond speech, we experiment with three common data sets. First, we do bird song detection \cite{stowell2019automatic} to solve a binary classification task with an objective to detect if the sound segment has a bird song in it. Finally, for music classification, we use the NSynth data set \cite{engel2017neural} of musical notes from different instruments.

More details about the data sets can be found in Table \ref{table:datasets}.

\section{Implementation Details}

\subsection{Augmentations}

As pointed out by \cite{zbontar2021barlow}, the Barlow Twins objective function is sensitive to the data augmentations used between the pair of samples for which the cross-correlation matrix is to be calculated and hence optimized. In this section, we will describe the different augmentations $K$ that is applied to both the twin batches before it is fed to our encoder network. We borrow ideas from \cite{niizumi2021byol} and implement a normalization step, followed by a mixup and random resized crop block each, which we will explain in detail in the following section.


\subsubsection{Normalization}

We normalize the mel-spectrogram, $x$, using:

\begin{equation}
\tilde{x}=\frac{x-\mu}{\nu},
\end{equation}

where $\mu$ and $\nu$ are the mean and standard deviation of the training samples, respectively.


\subsubsection{Mixup}
Mixup as an augmentation scheme has shown great success in supervised settings in CV \cite{zhang2018mixup}, were the first ones to have used it for an audio pre-training task in a SSL setting. They show significant performance boosts when they use it compared to when they do not (an average degradation of 8.4\% for 5 downstream tasks under the linear evaluation protocol).

The basic functionality of mixup is to \emph{mix} randomly selected input audio ($x_{k}$) in a small ratio from prior batches, which acts as background sound in the final \emph{mixed audio} ($\tilde{x}_{i}$). Formally, this would help the model learn embeddings invariant to noise.

\begin{figure}[!ht]
\centering
\includegraphics[width=0.8\textwidth]{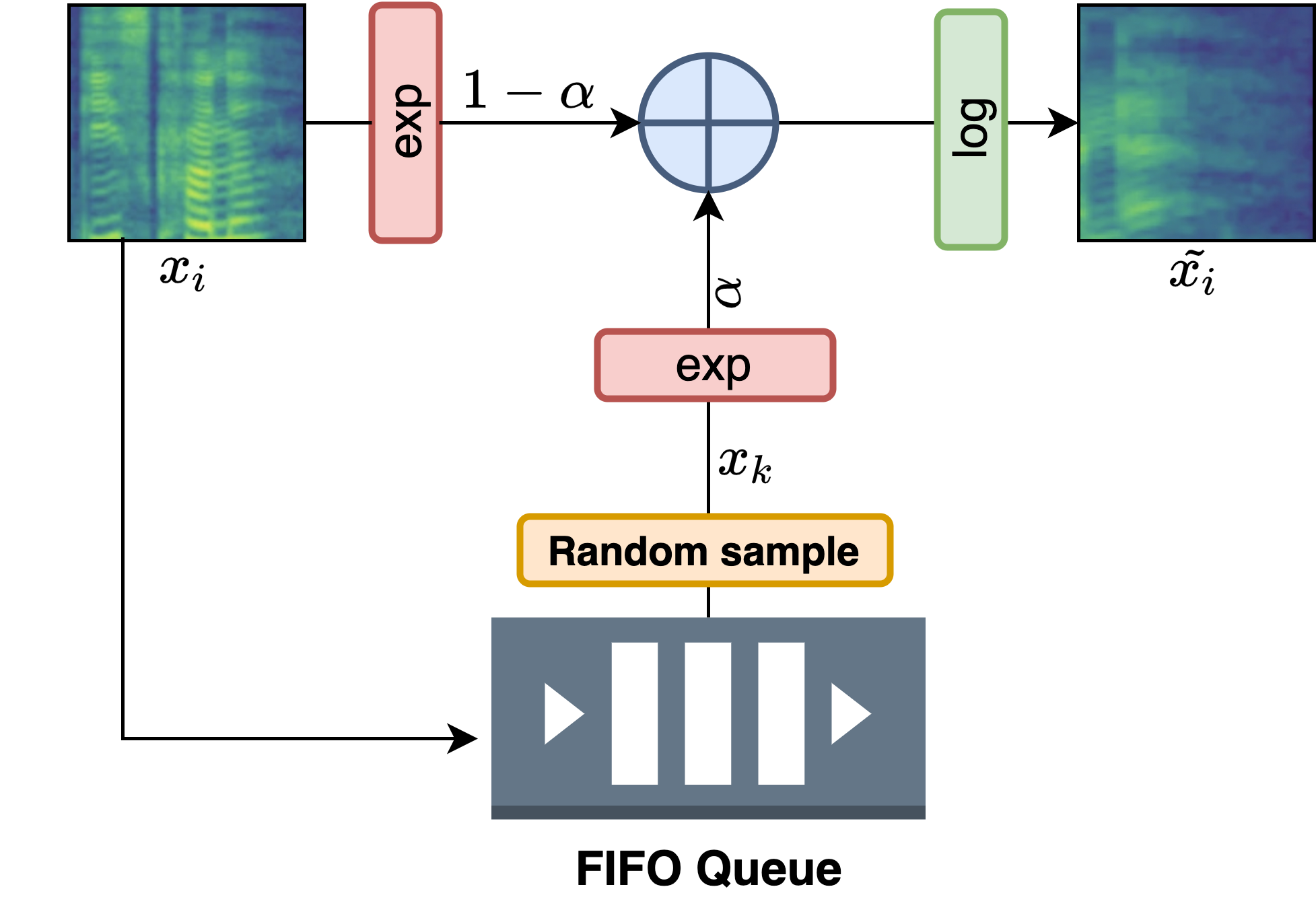} 
\caption{Mixup}
\label{fig:mixup}
\end{figure}

We use mixup in a similar setting to \cite{niizumi2021byol} where we use audio features only (instead of audio and labels both which was originally proposd by \cite{zhang2018mixup}), because of the lack of labels in a SSL setting. Since audio is log-scaled before passing it through this augmentation block, the input is converted to linear scale before mixup and converted back to log-scale again. A FIFO queue of size 2048 is maintained for storing past inputs, from where spectrograms are randomly sampled for mixup. 


\begin{figure}[!ht]
\centering
\includegraphics[width=0.8\textwidth]{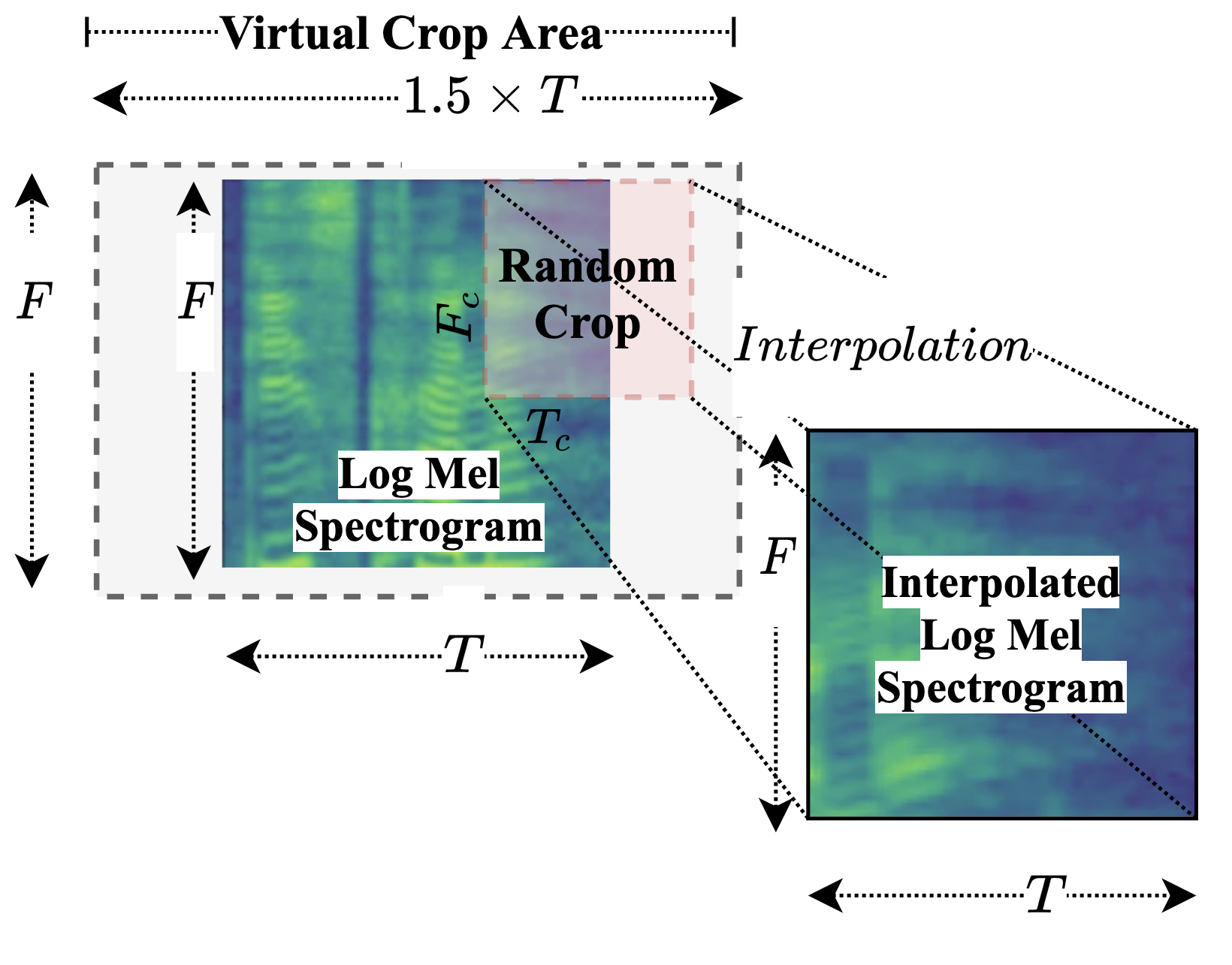} 
\caption{Random Resized Crop}
\label{fig:rcc}
\end{figure}

\subsubsection{Random Resized Crop}

Random resized crop (RRC) is one of the most prevalent image augmentation functions used in both supervised and semi-supervised settings in CV. Unlike mixup, RRC does not need labels. The primary motive of RRC is to crop a random portion of image and resize it to a given size. Most SSL algorithms use it \cite{chen2020simple,caron2021unsupervised,zbontar2021barlow} and especially \cite{zbontar2021barlow} show a significant drop in performance 

\begin{table*}[!ht]
    \centering
    \resizebox{17.5cm}{!}{%
    \begin{tabular}{|l|l|l|l|l|}
    \hline
    \textbf{Data Set} & \textbf{Target} & \textbf{No. of  Classes}  & \textbf{No. of  Samples} & \textbf{Avg.  Duration (sec)} \\
    \hline
    LibriSpeech (LBS)  & Speaker Identification &  585 & 28,538 & 12.69 \\
    \hline
    VoxCeleb 1 (VC)  & Speaker Identification &   1,211 & 153,397 & 8.20 \\
    \hline
    IEMOCAP (IC)  & Emotion Recognition &  4 & 4,490 & 4.49\\
    \hline
    Speech Commands V1 (SC-V1) & Keyword Recognition &   12 & 64,721 & 0.98 \\
    \hline
    Speech Commands V2 (SC-V2) & Keyword Recognition &   12/35 &  105,829 &0.98\\
    \hline
    Bird Song Detection (BSD) & Song detection   & 2 & 15,690 &10.08\\
    \hline
    VoxForge (VF) & Language Identification & 6 & 176,438 &6.68\\
    \hline
    NSynth (NS) & Musical Instruments Identification & 11 & 301,883  &4.00\\
    \hline
    \end{tabular}%
    }
    \caption{Dataset statistics for downstream benchmark tasks. The settings have been inspired from and is in-lines with prior-art \cite{saeed2020contrastive,niizumi2021byol}}
    \label{table:datasets}
\end{table*}

when they do not use it. In this paper, we use RRC in a fashion which tries to do a task exactly similar to the original RRC used in CV. For adapting it to the audio domain, we draw inspiration from \cite{niizumi2021byol}.

The input spectrogram consists of frequency bins $F$, and $T$ time frames. First, we sample the random crop area from a virtual crop boundary with
time frames longer than the input, i.e., $1.5\times T$. The size of the crop area is randomly sampled as:

\begin{equation}
    F_{c}=\left\lfloor\min \left(U\left(h_{1}, h_{2}\right), 1.0\right) \times F\right\rfloor
\end{equation}

\begin{equation}
T_{c}=\left\lfloor U\left(w_{1}, w_{2}\right) \times T\right\rfloor
\end{equation}

where $F_c$ and $T_c$ are the number of frequency bins and number of time frames of random crop size, respectively, $h_1$ and $h_2$ form a frequency bin range [$h_1$, $h_2$], $w_1$ and $w_2$ form a time frame range [$w_1$, $w_2$] and $U$ stands for the range function. We allow the dimension on the time axis to exceed the original dimension of the spectrogram through padding, but not on the frequency domain. Contents in the crop are then resized to the size of the original input with bicubic interpolation.

\subsection{Experimental Setup}

Given  an  audio  input  sequence,  we  extract  log-compressed mel-filterbanks with a window size of 64 ms, a hop size of 10 ms, and $N$ = 64 mel-spaced frequency bins in the range 60 – 7800  Hz. As mentioned earlier, for pre-training we randomly crop a small segment of the original audio from the AudioSet (originally 10 seconds each). This is in line with previous work in this domain \cite{saeed2020contrastive, niizumi2021byol}. For our pre-training setup, we use $T$ = 96 frames, which corresponds to 1024 seconds. For downstream tasks, the number of frames $T$ depends on the average duration of each downstream data set (for example 12.69 s for LibriSpeech and 4.49 s for IEMOCAP).

For our ConvNet encoder we choose an embedding size $h_x \in R^{2048}$. During pre-training, we pass $h_x$ through a set of linear layers $g$, also called the projection head, which contains two fully connected layers with 8192 units each and produces $g_x$ $\in$ $R^{8192}$, followed by a BatchNormalization layer, to make the embeddings zero mean and unit variance. For training and inference on downstream tasks, we discard $g$ after our pre-training step and replace it with a single linear layer with units equal to the number of classes in the task.


We pre-train our model using SSL for a total of 100 epochs. We learn the parameters of our model with the optimization scheme proposed in \cite{zbontar2021barlow, grill2020bootstrap}.  We use the LARS optimizer \cite{you2017large} with
a batch size of 1024. \cite{zbontar2021barlow} however show that Barlow Twins loss function performs well with smaller batches too.



As used in \cite{zbontar2021barlow}, we set the learning rate to 0.2 for the weights and 0.0048 for the biases. The learning rate is multiplied by the batch size and divided by 256. We use
a learning rate warm-up period of 10 epochs, after post which
we reduce the learning rate by a factor of 1000 and use a cosine decay schedule \cite{loshchilov2017sgdr} for the same. Finally, the trade-off parameter $\lambda$ of the loss function was set to 0.0051.

We train all our downstream tasks with Adam optimizer, a learning rate of $10^{-3}$, in batched-mode with a batch size of 64, and train it for a maximum of 100 epochs.

In this work, we have borrowed most of the hyper-parameter settings from the original implementation \cite{zbontar2021barlow}.



\section{Results}
\begin{table}[t]
\begin{tabular}{|l|l|l|l|}
\hline 
\textbf{Model} & \textbf{Params} & \textbf{Data} & \textbf{Epochs (SSL/Down.)}\\
\hline 
TRILL & 27M & 1M+ &$-$\\
\hline
COLA  & 11M & 2M & 500/- \\
\hline
BYOL-A  & 5.3M & 0.25M & 100/200  \\
\hline
DECAR & 11M & 0.2M & 100/50\\
\hline
SSAST & 87M & 2M & 8.5/50 \\
\hline
DeLoRes & \textbf{5.3M} & \textbf{0.25M} & 100/100\\
\hline
\end{tabular}
\caption{Comparison of DeLoRes with prior-art on general-purpose audio representation learning. DeLoRes is trained on a low-resource setting in terms of both data and compute compared to prior-art.}
\label{table:params}
\end{table}

\subsection{Linear Evaluation Protocol}

We compare the effectiveness of DeLoRes embeddings over 9 challenging and diverse downstream tasks. Table \ref{table:lineareval} reports a comparison with other SOTA methodologies on the \emph{linear evaluation protocol} where only the linear layer is trained for the particular downstream over DeLoRes embeddings. 

\begin{table*}[!ht]
\centering
\resizebox{18.5cm}{!}{%
\begin{tabular}{|l|l|l|l|l|l|l|l|l|l|}
\hline 
\textbf{Downstream Task} & \textbf{CBoW} & \textbf{SG} & \textbf{TemporalGap} & \textbf{Triplet Loss} & \textbf{TRILL} & \textbf{COLA} & \textbf{BYOL-A} & \textbf{DECAR} &  \textbf{DeLoRes}\\
\hline 
Speech Commands V1 & $-$ & $-$ & $-$ & $-$ & $7 4 . 0$ & $71.7$ & $-$ &$63.9$& $\mathbf{86.1}$\\
\hline
Speech Commands V2 (12) & $-$ & $-$ & $-$ & $-$ & $74.0$ & $-$ & $84.5$& $65.7$ &$\mathbf{85.4}$ \\
\hline
Speech commands V2 (35) & $30.0$ & $28.0$ & $23.0$ & $18.0$ & $-$ & $62.4$ & $\mathbf{87.2}$ & $-$ & $80.0$\\
\hline
LibriSpeech & $99.0$ & $\mathbf{1 0 0 . 0}$ & $97.0$ & $\mathbf{1 0 0 . 0}$ & $-$ & $\mathbf{1 0 0 . 0}$ & $-$ &$62.5$& $90.1$\\
\hline
VoxCeleb & $-$ & $-$ & $-$ & $-$ & $17.7$ & $2 9 . 9$& $31.0$ &$2.5$ & $\mathbf{31.2}$ \\
\hline
NSynth & $33.5$ & $34.4$ & $35.1$ & $25.7$ & $-$ & $63.4$ & $\mathbf{71.2}$ & $59.9$ & $66.3$\\
\hline
VoxForge & $-$ & $-$ & $-$ & $-$ & $\mathbf{8 8 . 1}$ & $71.3$ & $83.1$ & $46.0$ &$76.5$ \\
\hline
IEMOCAP & $-$ & $-$ & $-$ & $-$ & $-$ & $-$ & $-$ &$60.5$& $\mathbf{60.7}$ \\
\hline
Birdsong Detection & $71.0$ & $69.0$ & $71.0$ & $73.0$ & $-$ & $77.0$ & $-$& $76.4$ & $\mathbf{86.7}$\\
\hline

\end{tabular}%
}
\caption{Result comparison for \emph{linear evaluation protocol} setup. Results for approaches other than DeLoRes has been taken from literature. ``--" signifies that results were not reported for these tasks by these methods.}
\label{table:lineareval}
\end{table*}
\begin{table*}[!ht]
\centering
\begin{tabular}{|l|l|l|l|l|l|l|}
\hline 
\textbf{Downstream Task} & \textbf{TRILL} & \textbf{COLA} & \textbf{DECAR} & \textbf{Wav2Vec} & \textbf{SSAST} & \textbf{DeLoRes}\\
\hline 
Speech Commands V1 & $-$ & $\mathbf{98.1}$ & $97.6$ & $96.2$ & $96.2$ &$97.7$\\
\hline
Speech Commands V2 (12) & $91.2$ & $-$ & $97.6$ & $-$ & $-$ &$\mathbf{97.8}$ \\
\hline
Speech commands V2 (35)  & $-$ & $95.5$ & $-$ & $-$& $\mathbf{98.2}$ & $95.9$\\
\hline
LibriSpeech & $-$ & $\mathbf{1 0 0 . 0}$ & $97.0$ &$-$&$-$&$95.3$\\
\hline
VoxCeleb & $17.6$ & $37.7$ & $57.5$ & $56.6$ &$\mathbf{66.6}$ &$60.3$ \\
\hline
NSynth & $-$ & $73.0$ & ${78.4}$ & $-$&$-$ & $\mathbf{78.6}$\\
\hline
VoxForge & $94.1$ & $82.9$ & $76.5$ & $-$ & $-$ & $\mathbf{95.6}$ \\
\hline
IEMOCAP & $-$ & $-$ & $\mathbf{66.9}$ & $57.1$ & $59.8$ &$63.9$ \\
\hline
Birdsong Detection& $-$ & $80.2$& $\mathbf{90.3}$& $-$&$-$ & $\mathbf{90.3}$\\
\hline
\end{tabular}
\caption{Result comparison for \emph{transfer learning} setup. Results for approaches other than DeLoRes has been taken from literature. ``--" signifies that results were not reported for these tasks by these methods.}
\label{table:finetune}
\end{table*}

Even with our smaller architecture and lesser pre-training data, we are competitive to other prior-work in this domain, except BYOL-A which serves as the current SOTA \footnote{We only compare with prior-art which focuses on leaning \emph{general-purpose audio representations} and not all speech and audio SSL algorithms} for the \emph{linear evaluation protocol} for generalizing across 5 tasks as can be seen in Table \ref{table:lineareval}. BYOL-A was pre-trained using SSL on 10 times more data than our setting, with double the number of parameters to be learned (because of the presence of a teacher network) and was pre-trained for 400 epochs which is 4 times more than our implementation.

\subsection{Transfer Learning Setup}
Table \ref{table:finetune} reports a comparison with other SOTA methodologies on the \emph{transfer learning} setup where all the model weights are fine-tuned end-to-end on the downstream task. Though this would not a be fair comparison because a bigger model architecture might learn task-specific features better than ours, similar to the linear evaluation protocol, we show through our experiments that DeLoRes overpowers other approaches in the general-purpose audio representation learning space in 5 out of 9 tasks while losing marginally in the other 4 tasks.

\subsection{Low-Resource Setting}

Table \ref{table:params} shows the comparison of other SSL pre-training tasks with our proposed method, with respect to model size, training time (number of epochs) and total amount of data used for SSL (number of AudioSet samples). We claim our training to be on \emph{low-resource} for three main reasons:

\FloatBarrier
\begin{figure}[!ht]
\includegraphics[width=0.97\textwidth]{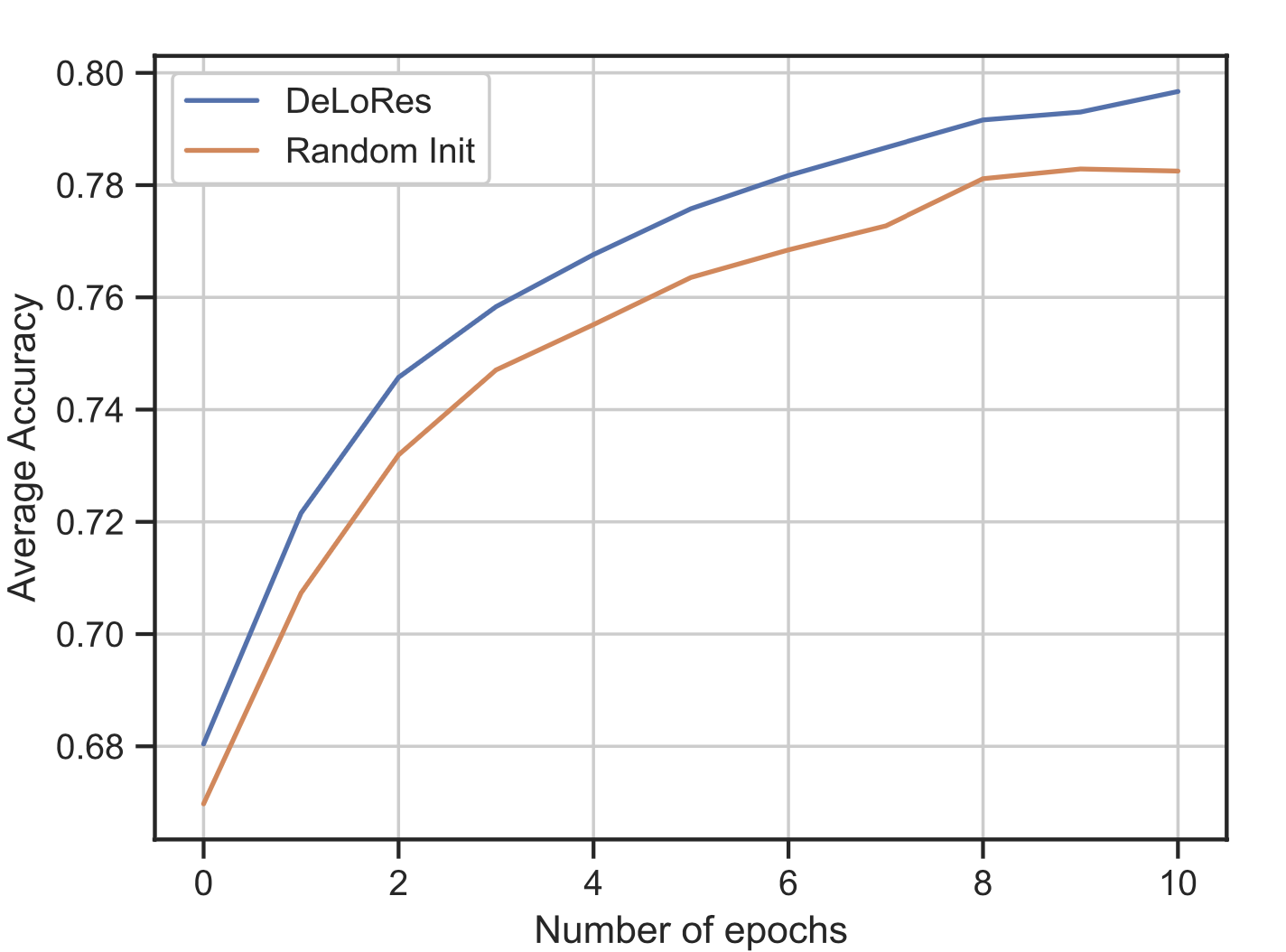}
\caption{Average score over 9 downstream tasks Vs Number of Epochs (11 epochs) in \emph{transfer learning setup}}
\label{fig:bird_compare}
\end{figure}
\FloatBarrier

\begin{itemize}
    \item We use only a fraction of the total pre-training data compared to prior work. SSL has been known to benefit from large amounts of data.
    \item Our architecture uses the least number of encoder parameters for SSL with unlabeled audio. BYOL-A, which uses the same network as ours, uses a separate teacher network which double the number of parameters needed for training. Fig. \ref{fig:transfer_vox} give a pictorial representation for the same for one particular downstream task.
    
\FloatBarrier
\begin{figure}[!ht]
\includegraphics[width=1.0\textwidth]{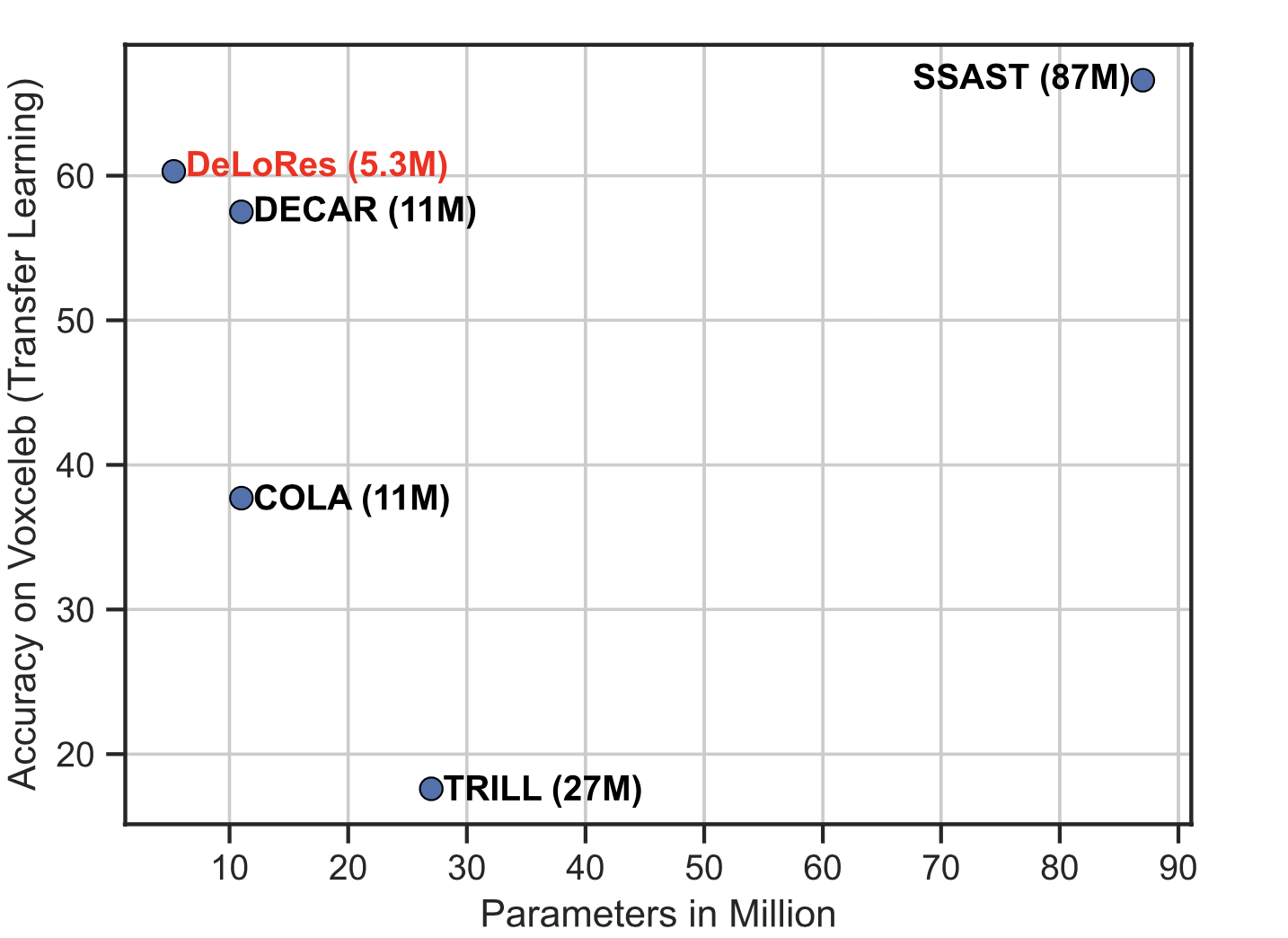} 
\caption{Number of parameters vs. performance for Voxceleb on the transfer learning setup}
\label{fig:transfer_vox}
\end{figure}
\FloatBarrier

     DeLoRes is competitve to prior art with a fraction of the total trainable parameters present in the model.
    \item We pre-train for much lesser number of epochs when compared to prior-art. Our model pre-training converges by 100 epochs.
\end{itemize}




\subsection{Analysis of Results}
As seen in Table \ref{tab:comp}, we show significant
gains compared to using DeLoRes embeddings as compared to randomly initialized embeddings. This makes
it very evident that we learn powerful representations from our pre-training step. The choice of tasks for Table \ref{tab:comp} is made such that all tasks for comparison have fewer training samples than our total pre-training data.

Additionally, we see in Fig. \ref{fig:bird_compare}, that our model starts at a much better accuracy on the first epoch, and jumps higher afterward in later epochs. With SSL based pre-training the model trains faster and better.

\FloatBarrier
\begin{table}[ht]
  \begin{tabular}{|l|l|l|l|l|}
    \hline
    \multirow{2}{*}{\textbf{Downstream Task}} &
      \multicolumn{2}{c}{\textbf{Random Init}} &
      \multicolumn{2}{c|}{\textbf{DeLoRes Init}} \\
     & freeze & fine & freeze & fine \\
    \hline
     SC-V1 & $52.9$ &$97.3$ & $86.1$& $97.7$ \\
     \hline
     SC-V2(12) & $53.3$ &$97.6$ & $85.4$& $97.8$ \\
     \hline
     SC-V2(35) & $54.7$ &$95.8$ & $80.0$& $95.8$ \\
     \hline
    LBS  &$57.7$ &$91.1$ & $90.1$& $95.3$ \\
    \hline
    VC  &$6.0$ &$55.2$ & $31.2$& $60.3$ \\
    \hline
    NS & $43.2$ &$77.7$ & $66.3$& $78.6$ \\
    \hline
    VF & $56.7$ &$94.4$ & $76.5$& $95.6$ \\
    \hline
    IC & $53.3$ &$60.3$ & $60.7$& $63.9$ \\
    \hline
    BSD &$60.3$ &$85.1$ & $86.7$& $90.3$ \\
    \hline
    \textbf{Avg} & $49.5$ & $80.7$ & $72.1$ & $83.0$ \\
    \hline
\end{tabular}
    \caption{Randomly initialized vs DeLoRes initialized performance comparison}
\label{tab:comp}
\end{table}
\FloatBarrier

\section{Conclusion}

In this paper, we propose a new approach to learning general-purpose audio representations via self-supervised learning using an invariance and redundancy-reduction framework. DeLoRes is simple by nature and avoids trivial solution by its nature of reconstruction. We also show that it is possible to learn good representations under a low-resource setting (both data and compute). DeLoRes is competitive with linear evaluation and does extremely well when fine-tuned end-to-end. Future work involves pre-training DeLoRes for more epochs to see it improves performance and analyze the effect of changing different hyperparameters and also the sensitivity of our approach to the different augmentations applied. In a future version of the paper we propose the LAPE Benchmark for a holistic evaluation of learned audio representations with 11 diverse downstream tasks, and DeLoRes-M, which gives state-of-the-art results on 9 out of the 11 tasks in LAPE.



\bibliography{aaai22}

\section{Acknowledgments}
We thank Centre For Development Of Advanced Computing (CDAC), as part of the National Language Translation Mission for providing us with the compute resources required for the experiments. Additionally we thank all other members of Speech Lab IITM and Verisk Analytics for their valuable feedback which helped us correct errors in our paper.

\end{document}